\documentclass{ws-procs975x65}  

\newcommand{\beq}{\begin{equation}}
\newcommand{\eeq}{\end{equation}}
\newcommand{\lag}{{\cal L}}

\begin{document}

\title{Electromagnetic shift arising from the Heisenberg-Euler dipole}

\author{Luca Bonetti\textsuperscript{1}, Santiago E. Perez Bergliaffa\textsuperscript{2}, and Alessandro D.A.M. Spallicci\textsuperscript{1,2}}

\address{
\textsuperscript{1}Universit\'e d'Orl\'eans\\ 
Observatoire des Sciences de l'Univers en r\'egion Centre, UMS 3116 \\
Collegium Sciences et Techniques, P\^ole de Physique\\
Centre Nationale de la Recherche Scientifique\\
Laboratoire de Physique et Chimie de l'Environnement et de l'Espace, UMR 7328\\
3A Av. de la Recherche Scientifique, 45071 Orl\'eans, France\\
\textsuperscript{2}Departamento de F\'{\i}sica Te\'orica, Instituto de F\'{\i}sica \\
Universidade do Estado do Rio de Janeiro\\
Rua S\~{a}o Francisco Xavier 524, Maracan\~{a}, 20550-900 Rio de Janeiro, Brasil
}

\begin{abstract}
We show that photons may be redshifted or blueshifted when interacting with the field of an overcritical dipole, which incorporates the one-loop QED corrections coming from vacuum polarization. Using the effective metric, it follows that such effect depends on the polarization of the photon. 
The shifts, plotted against the azimuthal angle for various values of the magnetic field, may show an intensity comparable to the gravitational redshift for a magnetar. To obtain these results, we have corrected previous literature.
\end{abstract}

\bodymatter

\section{Introduction}

Nonlinear electromagnetic (EM) theories have been widely studied 
both in the  classical and quantum realm. 
Born and Infeld's theory \cite{boin34}, in which the field of a charged particle is regular everywhere, is an example of the former type. Still on the classical side, the effect of gravity has been analyzed in several papers, including black holes sourced by a nonlinear EM field \cite{brpb14}. The latter type is 
based on the nonlinear corrections that Maxwell's Lagrangian gains due to vacuum polarization, as shown by Heisenberg 
and Euler (HE)\cite{heeu36}.
The corrections are important when the fields reach a value comparable to the critical 
electric  $E_c \approx 1.3\times 10^{18}$ V/m or
magnetic $B_c \approx 4.4\times 10^{13}$ G field. 
There are several testable predictions that follow from the Heisenberg-Euler Lagrangian: the Schwinger effect \cite{sc51} 
,the birefringence of the vacuum under a strong magnetic field
 \cite{bbbb70},  
currently under experimental observation 
\cite{zaetal13,befobari12}, and photon splitting \cite{adbacaro70} in a laser field
\cite{dpmike07}.

Beyond the experimental effort on earth, some compact astrophysical objects may offer the possibility of testing the quantum vacuum. The magnetars 
are neutron stars endowed with an overcritical magnetic field \cite{me08}, with estimated values up to
$10^2B_c$ at their surface \cite{me13}. Several consequences of such intense field have been studied: 
such as 
the lensing of the light emitted
by background astronomical objects due 
to the optical properties of quantum vacuum in the presence of a magnetic field \cite{duroribi05},
the polarization phase lags due to the index of refraction of the vacuum \cite{hesh00}, 
and the influence of the quantum vacuum friction on the spindown of pulsars \cite{duribi12}.

Many of these effects are related to photon propagation in a background field, which can be studied using the effective metric. 
The propagation of the high energy excitations of a nonlinear EM theory on a fixed 
background is governed by an effective metric \cite{agdupl81,nopbsa00} that depends on the background spacetime metric, on the background EM field configuration, and on the details
of the nonlinear dynamics obeyed by the EM field \cite{balivi05}. There are numerous applications of the effective metric in pure gravity
\cite{nopbsa00}
, and cosmology \cite{nogosapb07}. 
In an astrophysical setting, it was shown for the case of BI's theory \cite{mcsa04a} and for the low-field correction to Maxwell's Lagrangian obtained from HE's Lagrangian \cite{mcsa04b} that 
high energy photons coming from a compact object endowed with an overcritical field would display an EM redshift, in addition to the 
gravitational redshift. The EM redshift (or blueshift, as we shall see below) is rooted in the nonlinearities of the Lagrangian, hence it is absent in Maxwell's theory. 
Here we investigate the electromagnetic shift (EMS) of photons in a dipole field which is a solution of the full QED one-loop  corrected Lagrangian, in a flat space-time background \cite{hehe97}, through the effective metric.

\section{Effective metric}
How the propagation of high energy perturbations on a fixed background of a nonlinear theory with more that one degree of freedom may display birefringence and/or bimetricity was originally discussed 
\cite{agdupl81,nopbsa00,balivi05} for the EM field. A general Lagrangian ${\cal L}$ for a nonlinear EM field
can be an arbitrary function of the invariants $F$ and $G$, given by 
$F=\frac{1}{4}F_{\mu\nu}F^{\mu\nu}$, and $G=\frac{1}{4}^*F_{\mu\nu}F^{\mu\nu}$, where
$^*F_{\mu\nu}=\frac{1}{2} \epsilon^{\mu\nu\lambda\sigma}F_{\lambda\sigma} $.   
The equations of motion following from 
such a Lagrangian
in a flat background are \cite{nopbsa00}
\begin{equation}
\left[\sqrt{-\eta}\left({\cal L}_FF^{\mu\nu}+{\cal L}_{^*F}\;\;^*F^{\mu\nu}\right)\right]_{;\nu}=0,
\end{equation}
where $\eta$ is the determinant of the background metric $\eta_{\mu\nu}$; the covariant derivative is constructed with $\eta_{\mu\nu}$, and ${\cal L}_X \equiv \partial{\cal L}/\partial X$ \cite{nopbsa00}.
The fields also have to obey the identity $F_{[\mu\nu ,\lambda]}=0$.
By perturbing the equation of motion with respect to a fixed background EM field, keeping only terms linear in the perturbation, and applying the eikonal approximation \cite{vibali02}), it follows that in the high energy limit the propagation of each of the two polarizations of the EM field is governed by an effective metric, given by
\begin{equation}
^{(1)}\widetilde g^{\mu\nu} = \left[\lag_{F}\,\eta^{\mu\nu}  - 
4\, \lag_{FF} \,{F^{\mu}}_{\alpha} \,F^{\alpha\nu}\right]_0,
\label{geffecplus}
\end{equation}
\begin{equation}
^{(2)}\widetilde g^{\mu\nu} = \left[\left(\lag_{F}-2F\lag_{GG}\right)\,\eta^{\mu\nu}  - 
4\, \lag_{GG} \,{F^{\mu}}_{\alpha} \,F^{\alpha\nu}\right]_0,
\label{geffecminus}
\end{equation}
where the subscript zero means that all the quantities in both effective metrics are evaluated at the background field. 

$$
1+ z = \frac{(\widetilde g^{00})^{-1}(E)}{(\widetilde g^{00})^{-1}(R)}.
$$
%
The shift can be computed using the usual general relativity formulas, keeping into account that the photons, in these theories, are accelerated by the nonlinearities and therefore, do not move along the geodesics of the background metric\cite{agdupl81,nopbsa00}. Hence one has
\beq
^{(1)}z = \lag_F(r_E)-1,
\label{r1}
\eeq
and 
\beq
^{(2)}z = \left[   
\lag_F(r_E)-2 F(r_E) \lag_{G G}(r_E) \right] -1  .
\label{r2}
\eeq
where the index $E$ ($R$) refers to the emission (reception) point.
In both expressions, the reception point was taken at an infinite distance from $E$, where the effective metric reduces to that of Minkowski.

These expressions are valid for any nonlinear EM theory. In the next section, we shall 
work with the HE Lagrangian.

\section{Heisenberg-Euler Lagrangian}

 The Heisenber-Euler Lagrangian is an effective Lagrangian for the one-loop QED. Its
 form is given in Lundin's paper \cite{lu09} Eq. (1).
This Lagrangian is valid in the so-called soft photon
approximation ($\omega \ll $ m). \cite{lu09}  
In case of a zero electric field, 
$F=B^2/2$, $G=0$. 
It is worth pointing out that although the HE Lagrangian was originally derived under the assumption of a constant electric or magnetic field, it can still be used in situations in which the fields are inhomogeneous, through a derivative expansion \cite{dunne04}. In this case, the lowest-order approximation is given by the HE Lagrangian evaluated at the inhomogeneous field. As seen from Eqns. (\ref{r1}) and (\ref{r2}), the dependence of ${\cal L}_F$ and ${\cal L}_{GG}$ with the field is needed
to calculate the EMS. In terms of $\xi = B/B_c$, such functions are given \cite{lu09}  (see Eq. 3a-3c).


\section{Calculation of the electromagnetic shift}

\begin{figure}
\includegraphics[width=2.3in]{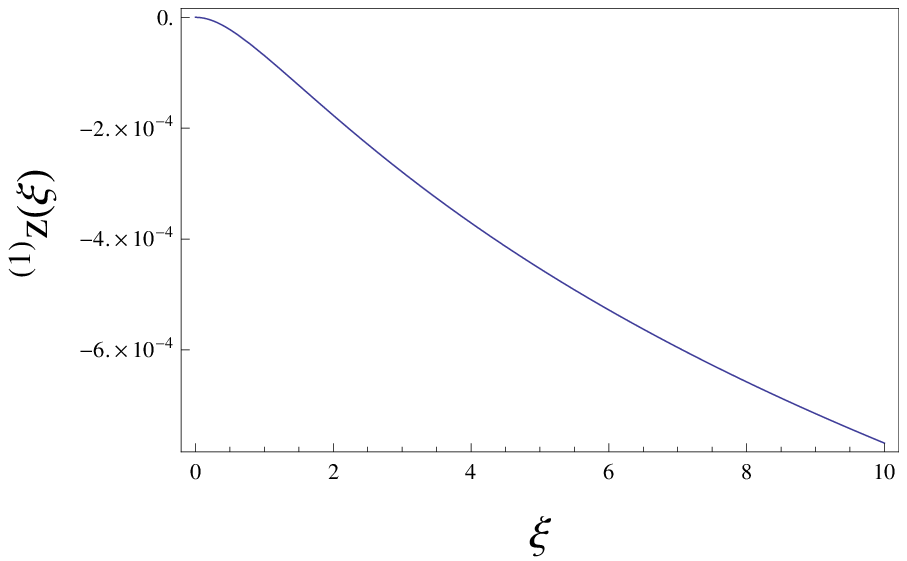}
\quad
\includegraphics[width=2.3in]{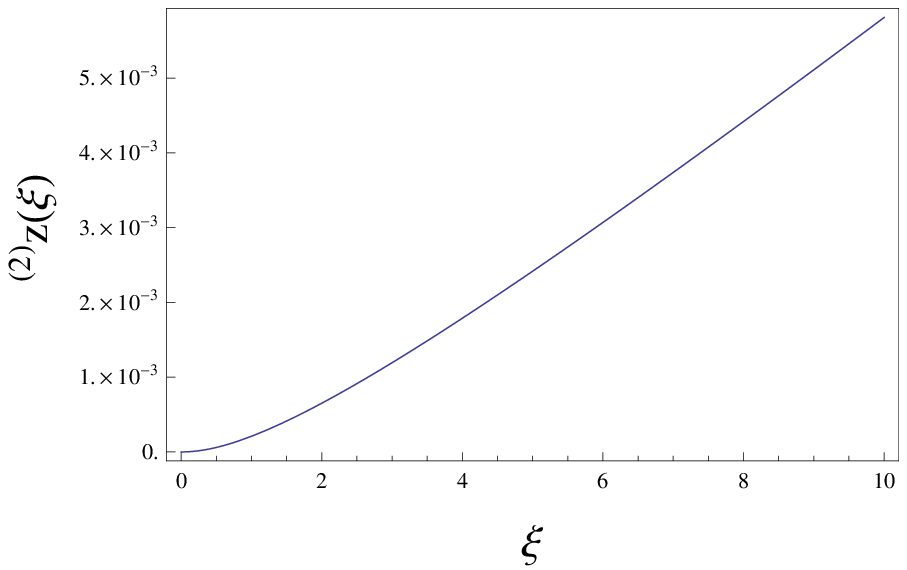}
\caption{The figure on the left displays the dependence of $^{(1)}z$ with $\xi$,
while in that on the right, the dependence of $^{(2)}z$ with $\xi$ is shown.}
\label{zxi}
\end{figure}

To compute the EMS associated to each polarization, we use as the background solution that of a dipole 
oriented along the $z$ axis,
namely
\begin{equation}
\phi_0({\bf r}) = \sqrt{\frac{4 \pi}{3}} \frac{d}{r^2}
Y_{10}(\theta,\phi),
\label{phi0sph}
\end{equation}
where the background field is calculated from the potential using $\mathbf{B_0} = -\boldsymbol{\nabla} \phi_0$,
and $d$ is the dipolar moment.
Since the relevant expressions in both effective metrics are already $O(\alpha )$, only the zero-order solution for the field
is needed to calculate the EMS
\footnote{The $O(\alpha)$ correction for the dipole has been calculated \cite{hehe97}.}
Fig.~\ref{zxi} presents the dependence of the EMS with the variable $\xi$ for both polarizations. 
The figures show that the EMS is small in both cases 
as expected from an effect arising from a quantum correction. It is important 
to remark that the EMS 
is negative for the first polarization (hence a redshift)
and positive (so it is actually a blueshift) and larger
for the second. 

\begin{figure}
\includegraphics[width=2.3in]{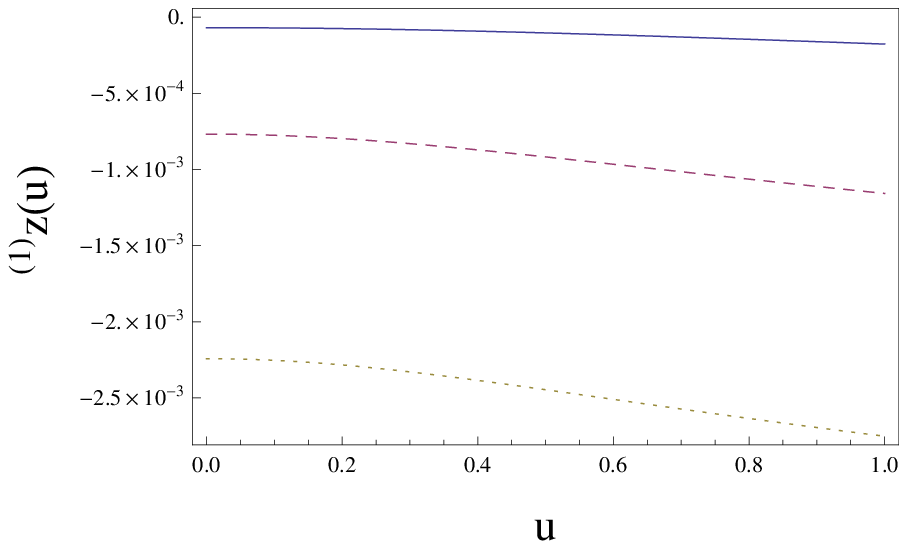} \quad
\includegraphics[width=2.3in]{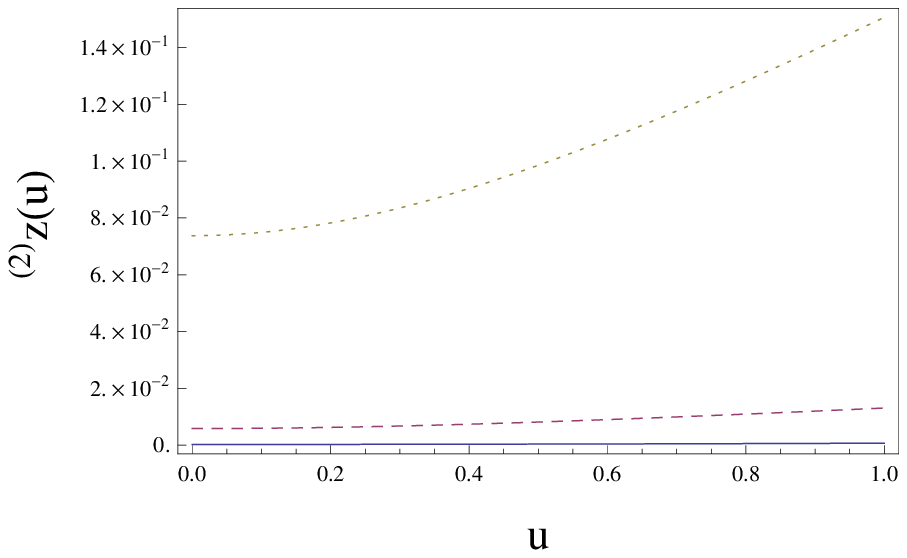}
\caption{Plot of $^{(1)}z$ (left) and $^{(2)}z$ in terms of $u=\cos\theta$, 
for $\lambda = 1,10,100$ (continuous, dashed, and dotted line respectively). In the plot on the left, the EMS is negative and decreases with the field. The opposite happens in the plot on the right.}
\label{zu}
\end{figure}

With the introduction of the parameter $\lambda$, given by the quotient of the mean dipole background field at $r=r_E$ and the critical field,
$$
\lambda=\frac{d}{r_E^3B_c}, 
$$
it follows that
$$
\xi = \frac{B_0}{B_c} = \lambda (1+3\cos^2\theta). 
$$
Hence, the EMS given by Eqns. (\ref{r1}) and (\ref{r2}) is a function of $\theta$ for 
each value of $\lambda$.
Fig.~\ref{zu} shows the dependence with $\cos\theta$ of the EMS for three values of 
$\lambda$, which are relevant for magnetars \cite{me08}). 
The plots show that 
in both cases the EMS increases (in absolute value) with the field, but while $^{(1)}z$ is maximum at the equator and minimum at the poles, $^{(2)}z$ behaves in the opposite way.
It also follows from the plots 
that the EMS is typically one order of magnitude larger for the second polarization.  

\section{Conclusions}

We have derived and computed the EMS that photons experience when emitted at a given point of a 
dipole field with QED one-loop corrections and travel to infinity. 
The EMS evidences a polarization dependent effect.  
The energy exchange between photons and the background field is due to their interaction through the nonlinearities of the theory (since the EMS is null when $\alpha = 0$). 

This new effect may be important at least in two different areas. In the astrophysics of magnetars,
gravitation and rotation should not be neglected, but the significance of the EMS can be seen in a simpler setting, taking into account that
the EMS adds to the gravitational redshift \cite{mcsa04a}. 
For a star of two solar masses and a radius of $10$ km (typical values for a magnetar), the gravitational redshift (calculated using the Schwarzschild-Droste metric, see Ref.~ \refcite{ro02}) would be approximately 0.2. Hence, as shown in Figure \ref{zu}, for strong enough background fields, the EMS can reach values that are of the order of the gravitational redshift (for one of the polarizations).
This effect would lead to a variation in the shift of photons coming from the surface of the star and hence in the ratio of the mass and the radius. Our results may be relevant also in experimental settings to detect the signature of the QED vacuum, in particular those that use an overcritical dipole magnetic field, such as PVLAS and BMV \cite{zaetal13,befobari12}. 
We hope to report on these developments in future publications.

\bibliographystyle{ieeetr}

\end{document}